\documentclass[11pt,twocolumn,oneside]{IEEEtran} 
\newcommand{\jfpackage}[1] {\usepackage{def/#1}}
\newcommand{\indef}[1] {\input{./def/#1}}
\jfpackage{jf-all}
\usepackage{xcolor}
\newcommand{\colorr}[1] {#1}

\usepackage{upgreek} 
\usepackage{hyperref}
\usepackage{cite}
\newcommand{\codo}[1] {}

\ifx\TScase\undefined
\def\TScase{teach} 
\fi

\newcommand{\ifStudy}[1] {}
\newcommand{\ifTeach}[1] {}
\ifthenelse{\equal{\TScase}{study}}{\renewcommand{\ifStudy}[1]{#1}}{}
\ifthenelse{\equal{\TScase}{teach}}{\renewcommand{\ifTeach}[1]{#1}}{}

\ifTeach{
\newcommand{\citl}[1] {\cite{#1}} 
\newcommand{\citesl}[2] {\cite{#1#2}} 
\newcommand{\citenew}[1] {\cite{#1}} 
\long\def\ifshort#1 {} 
\long\def\iflong#1 {#1} 
\newcommand{\add}[1] {#1} 
}

\ifStudy{
\newcommand{\citl}[1] {} 
\newcommand{\citesl}[2] {\cite{#1}} 
\newcommand{\citenew}[1] {\add{\cite{#1}}} 
\long\def\ifshort#1 {#1}
\long\def\iflong#1 {}
\newcommand{\add}[1] {\textcolor{blue}{#1}} 
}

\newcommand{\wiki}[1]	{\href{#1}{[wiki]}}

\newcommand{\term}[1]	{\index{#1}\textbf{\colorr{#1}}}

\newcommand{\tofref}[1]	{\todoa{Fig. (XXX)}}

\newcommand{\todof}[1]	{\todoa{Fig. (XXX)}}

\renewcommand{\defequ}	{\xmath{\triangleq}}

\newcommand{\arrow}	{\inmath{\rightarrow}}

\newcommand{\nonum}	{\nonumber}

\newcommand{\cset}[2]	{\braces{#1 \mbox{ \textbf{\large $:$} } #2}}

\newcommand{\vveps}	{\blmath{\varepsilon}}
\newcommand{\veta}	{\blmath{\eta}}
\newcommand{\vgam}	{\blmath{\gamma}}

\newcommand{\vh}	{\blmath{h}}

\newcommand{\vs}	{\jfunop{\bm{s}}}

\newcommand{\w}	{\blmath{w}}

\newcommand{\subrm}[1]	{_{\mathrm{#1}}}

\newcommand{\bset}[1]	{\xmath{\braces{#1}}}

\renewcommand{\reals}	{\jfunl{\mathbb{R}}}
\newcommand{\complex}	{\jfunl{\mathbb{C}}}

\newcommand{\vzero}	{\bmath{0}}

\newcommand{\bOm}	{\blmath{\Omega}}

\newcommand{\A}	{\blmath{A}}
\newcommand{\B}	{\blmath{B}}
\newcommand{\C}	{\blmath{C}}
\newcommand{\D}	{\blmath{D}}

\newcommand{\F}	{\blmath{\mathsf{F}}}
\newcommand{\G}	{\blmath{G}}
\renewcommand{\H}	{\blmath{H}}
\newcommand{\I}	{\blmath{I}}

\renewcommand{\P}	{\blmath{P}}

\newcommand{\T}	{\blmath{T}}

\newcommand{\X}	{\blmath{X}}
\newcommand{\Xh}	{\X^`}
\newcommand{\Y}	{\blmath{Y}}

\newcommand{\norminf}[1]	{\xmath{\norm{#1}_{\infty}}}

\newcommand{\vnorm}	{\norm}

\newcommand{\normi}[1]	{\xmath{\vnorm{#1}_1}}

\newcommand{\normii}[1]	{\xmath{\vnorm{#1}_2}}

\newcommand{\mnorm}[1]	{\xmath{|\!|\!| #1 |\!|\!|}}

\newcommand{\mnormfrob}[1]	{\xmath{\mnorm{#1}\subrm{Frob}}}

\newcommand{\mnormii}[1]	{\xmath{\mnorm{#1}_2}}

\newcommand{\reg}	{\jfunl{\upbeta}}

\newcommand{\px}	{\jfunlv{\mathrm{x}}}

\newcommand{\jcal}[1]	{\jfunl{\mathcal{#1}}}

\newcommand{\cD}	{\jcal{D}}

\newcommand{\zh}	{\xmath{\hat{\z}}}

\newcommand{\cZ}	{\jcal{Z}}

\newcommand{\vu}	{\jfunop{\bm{u}}}
\newcommand{\vv}	{\jfunop{\bm{v}}}
\newcommand{\y}	{\blmath{y}}

\newcommand{\xsym}	{x}

\newcommand{\x}	{\blmath{\xsym}}

\newcommand{\xh}	{\jfunop{\bm{\xsym}}^`}

\newcommand{\z}	{\blmath{z}}

\newcommand{\nLsym}	{\ooalign{$\mathsf{L}$\cr\kern+0.0em\textsf{-}}}

\newcommand{\potsym}	{\inmath{\psi}}
\newcommand{\pot}	{\jfunop{\potsym}}

\newcommand{\kostsym}	{\inmath{\Psi}}
\newcommand{\kost}	{\jfunop{\kostsym}}

\newcommand{\Kx}	{\jfunbp{\kostsym}{\x}}

\newcommand{\vd}	{\blmath{d}}

\newcommand{\mleq}	{\xmath{\preceq}}

\renewcommand{\F} {\blmath{F}}
\newcommand{\pu} {\jfunl{\vec{\nu}}} 
\newcommand{\Lips} {\jfunl{\mathcal{L}}}

\title{Optimization methods for MR image reconstruction
\iflong{\\(Long version)}
}

\author{Jeffrey~A.~Fessler,~\IEEEmembership{Fellow,~IEEE}%
\thanks{%
J.~A.~Fessler (fessler@umich.edu)
is with the 
EECS Department,
Univ. of Michigan.
Research supported in part by
NIH Grants
R01 EB023618,
U01 EB026977,
and
R21 AG061839.
\par
For IEEE SPMag special issue on
``Computational MRI: Compressed Sensing and Beyond.''
\par
\ifshort{
This is an abbreviated version to fit the SPMag page limits.
See
\url{http://arxiv.org/abs/1903.03510}
for a longer version with more text and references.
}
\iflong{
This is the longer arXiv version
that has more text and references.
}
} 
} 

\newcommand{\uncite}[1]{\relax}

\begin{document}

\maketitle

\ifshort{\vspace*{-3em}}

{
\begin{abstract}%
%
The development of compressed sensing methods%
\citl{candes:08:ait}
for \add{magnetic resonance} (MR) image reconstruction%
\citl{lustig:08:csm}
led to an explosion
of research
on models
and optimization algorithms
for MR \add{imaging} (MRI).
\add{
Roughly 10 years after such methods
first appeared in the MRI literature%
\citl{block:07:urm},
the U.S.~\add{Food and Drug Administration} (FDA) approved
certain compressed sensing methods
for commercial use%
\citl{fda:17:ge,fda:17:siemens,fda:18:philips},
making compressed sensing
a clinical success story for MRI.
}
This review paper summarizes
several key models and optimization algorithms
for MR image reconstruction,
including both the type of methods
that have FDA approval for clinical use,
as well as more recent methods
being considered in the research community
that use data-adaptive regularizers.
\comment{
One impetus for this paper
is that ``off the shelf'' optimization methods
have rarely been the best choice
for solving optimization problems in MR image reconstruction,
due to the large volume of MRI data
collected by clinical systems
and practical time constraints on processing time.
Instead,
}
\add{Many}
algorithms
have been devised
that exploit the structure of the system model
and regularizers used in MRI;
this paper strives
to collect such algorithms
in a single survey.
\iflong{
Many of the ideas used
in optimization methods for MRI
are also useful
for solving other inverse problems.
}

\comment{
My timetable will follow that of the special issue,
with submission of a full article by Mar.~1, 2019.
As a side note,
I am teaching a special topics course
on optimization methods for signal and image processing
during the W19 semester,
and there is considerable synergy
between the course topics and this paper.
Also, I will be presenting
a closely related tutorial course at ISBI 2019
and I hope to use this paper
as reference material for that course.
}

\end{abstract}
}

\section{Introduction}
\label{sec:intro}

\subsection{Scope}

Although the paper title begins with ``optimization methods,''
in practice one first defines a model and cost function,
and then applies an optimization algorithm.
There are several ways
to partition the space
of models, cost functions and optimization methods
for MRI reconstruction,
such as:
smooth vs non-smooth cost functions,
static vs dynamic problems,
single-coil vs multiple-coil data.
This paper focuses
on the \emph{static} reconstruction problem
because the dynamic case
is rich enough
to merit its own survey paper%
\citl{christodoulou::adm}.
This paper emphasizes algorithms
for multiple-coil data
(parallel MRI%
\citl{roemer:90:tnp,hamilton:17:rai})
because modern systems all have multiple channels
and advanced reconstruction methods
with under-sampling
are most likely to be used for parallel MRI scans.
Main families of parallel MRI methods include
``SENSE'' methods
that model the coil sensitivities in the image domain
\citesl{pruessmann:99:sse}{,pruessmann:01:ais},
``GRAPPA'' methods
that model the effect of coil sensitivity in k-space
\cite{griswold:02:gap},
and ``calibrationless'' methods
that use low-rank \add{or joint sparsity} properties
\citesl{shin:14:cpi}{,trzasko:11:cpm,majumdar:12:clm,balachandrasekaran:18:cfb}.
This paper considers all three approaches,
emphasizing SENSE methods for simplicity.%
\footnote{%
%
Jupyter notebooks with code in the open source language Julia
\citl{bezanson:17:jaf}
\add{that reproduce the figures in this paper
are available
in the Michigan Image Reconstruction Toolbox (MIRT)
at}
\url{http://github.com/JeffFessler/MIRT.jl}
%

}

\subsection{Measurement model}

The signals recorded by the sensors (receive coils)
in MR scanners
are 
linear functions
of the object's transverse magnetization.
That magnetization
is a complicated and highly nonlinear function
of the RF pulses,
gradient waveforms,
and tissue properties,
governed by the physics of the Bloch equation
\citesl{wright:97:mri}{,fessler:10:mbi,doneva::aoo}.
Quantifying tissue properties
using nonlinear models
is a rich topic of its own%
\citenew{ma:13:mrf}%
\citl{nataraj:18:dfm,cheng:12:pma,zhao:14:mbm,mehta:19:mrf},
but we focus here
on the problem
of reconstructing images of the transverse magnetization
from MR measurements.

Ignoring noise,
a vector $\vs \in \complex^M$
of signal samples
recorded by a MR receive coil
is related (typically)
to a discretized version
$\x \in \complex^N$
of the transverse magnetization
via a linear Fourier relationship:
\be
\vs = \F \x
,\quad
F_{ij} = \expni{2\pi \pu_i \cdot \px_j}
,\ \mystack{i=1,\ldots,M \\ j=1,\ldots,N,}
\ee{e,s=Fx}
where \pu_i denotes the k-space sample location
of the $i$th sample
(units cycles/cm)
and \px_j denotes the spatial coordinates
of the center of the $j$th pixel
(units cm).
In the usual case where the pixel coordinates
\bset{\px_j}
and k-space sample locations
\bset{\pu_i}
are both on appropriate Cartesian grids,
matrix \F
is square
corresponds to the (2D or 3D) discrete Fourier transform (DFT).
In this case
$\F\inv = \rat 1/N \F'$
so reconstructing \x from \vs
is simply an inverse \add{fast Fourier transform} (FFT),
and that approach is used
in many clinical MR scans.

The reconstruction problem becomes more interesting
when the k-space sample locations
are on a non-Cartesian grid%
\citesl{pruessmann:01:ais}{,fessler:03:nff},
when the scan is ``accelerated''
by recording $M < N$ samples,
when non-Fourier effects like magnetic field inhomogeneity are considered
\cite{sutton:03:fii}
and/or when there are multiple receive coils.
In parallel MRI,
let \vs_l denote the samples
recorded by the $l$th of of $L$ receive coils.
Then one replaces the model
\eref{e,s=Fx}
with
\be
\vs_l = \F \C_l \x
,\ee{e,s=FCx}
where \C_l is a $N \times N$ diagonal matrix
containing the coil sensitivity pattern
of the $l$ coil on its diagonal.
Note that \F does \emph{not} depend on $l$;
all coils see the same k-space sampling pattern.
Stacking up the measurements
from all coils
and accounting for noise
yields
the following basic forward model in MRI:
\be
\mymat{\y_1 \\ \vdots \\ \y_L}
=
\y = \undertext{(\I_L \otimes \F) \C}{\A} \x + \vveps
,\
\C = \mymat{\C_1 \\ \vdots \\ \C_L}
,\ee{e,y=Ax}
where
$\A \in \complex^{ML \times N}$
denotes the system matrix,
$\y \in \complex^{ML}$
denotes the measured k-space data,
and
$\x \in \complex^N$
denotes the latent image.
The noise in k-space is well modeled
as complex white Gaussian noise%
\citl{macovski:96:nim}.
For extensions that consider
other physics effects
like relaxation and field inhomogeneity,
see
\cite{fessler:10:mbi}.

The goal in MR image reconstruction
is to recover \x from \y
using the model \eref{e,y=Ax}.
All MR image reconstruction problems
are under-determined
because the magnetization of the underlying object
being scanned
is a space-limited continuous-space function on $\reals^3$,
yet only a finite number of samples are recorded.
Nevertheless,
the convention in MRI is to treat the object
as a finite-dimensional vector
$\x \in \complex^N$
for which
$M \geq N$ appropriate Cartesian k-space samples
is considered ``fully sampled''
and any $M < N$ is considered
\add{``accelerated.''}
\iflong{
The term ``compressed sensing''
\cite{ye:19:csm} 
in this setting
might simply mean
that the k-space sampling is
\A is a wide matrix,
\ie, $LM < N$,
or might imply
that the sampling pattern
satisfies some sufficient condition
for ensuring good recovery of \x from \y.
}
Sampling pattern design
is a topic of ongoing interest
\citl{adcock:13:bcr,adcock:15:gsa,adcock:17:btc},
with renewed interest
in data-driven methods
\citesl{baldassarre:16:lbc}{,cao:93:frm,gozcu:18:lbc}.
\iflong{
One MRI vendor uses a spiral phyllotaxis sampling pattern
for 3D imaging
\cite{forman:14:hr3}
that emphasizes
the center of k-space.
}

The matrix \F
in \eref{e,y=Ax}
is known prior to the scan,
because
the k-space sample locations
\bset{\pu_i}
are controlled by the pulse sequence designer.
\iflong{
(Calibration methods are sometimes needed
for complicated k-space sampling patterns
\cite{duyn:98:scm}.)
}
In contrast,
the coil sensitivity maps
\bset{\C_l}
depend on the exact configuration of the receive coils
for each patient.
To use the model
\eref{e,y=Ax},
one must determine
the sensitivity maps
from some patient-specific calibration data,
\eg,
by joint estimation
\citesl{ying:07:jir}{,uecker:08:irb,ma:15:isi,she:15:irf,irfan:16:sme},
regularization%
\citenew{block:07:urm}%
\citl{allison:13:are},
or subspace methods
\cite{uecker:14:eae}. 

\section{Cost functions and algorithms}

\subsection{Quadratic problems}
When $ML \geq N$,
\ie,
when the total number of k-space samples
acquired across all coils
exceeds the number of unknown image pixel values,
the linear model
\eref{e,y=Ax}
is over-determined.
\add{If additionally \A is well conditioned,
which depends on the sampling pattern
and coil sensitivity maps,
then}
it is reasonable
to consider
an ordinary least-squares estimator%
\iflong{
\footnote{%
Coil coupling induces noise correlation between coils
that one should first whiten
\cite{ramani:11:pmi}.
Often the data from multiple coils
is condensed to a smaller number of virtual coils
to save computation and memory
\cite{buehrer:07:acf}.
}
}
\dena{
\xh &= \argmin{\x \in \complex^N}
\Half \normii{\A \x - \y}^2
= (\A'\A)\inv \A'\y
\nonum
\\&
= \paren{\textstyle \sum_{l=1}^L \C_l' \F' \F \C_l}\inv
\paren{\textstyle \sum_{l=1}^L \C_l' \F' \y}
.}{e,xh,ls}
In particular,
for fully sampled Cartesian k-space data
where
\(
\F\inv = \rat1/N \F'
,\)
this least-squares solution simplifies
to
\(
\xh
= \paren{\sum_{l=1}^L \C_l' \C_l}\inv
\paren{\sum_{l=1}^L \C_l' \F\inv \y}
,\)
which is trivial to implement
because each \C_l is diagonal.
This is known as the optimal coil combination approach
\cite{roemer:90:tnp}.
For regularly under-sampled Cartesian data,
where only every $n$th row of k-space is collected,
the matrix $\F'\F$ has a simple block structure
with $n \times n$ blocks
that facilitates non-iterative block-wise computation
known as SENSE reconstruction
\cite{pruessmann:99:sse}.
This form of least-squares estimation
is used widely in clinical MR systems.

\subsection{Regularized least-squares}

For under-sampled problems
$(ML < N)$
the LS solution 
\eref{e,xh,ls} is not unique.
\add{Furthermore,
even when $ML \geq N$
often \A is poorly conditioned,
particularly for non-Cartesian sampling.
}
Some form of regularization is needed
in \add{such} cases.
Some early
MRI reconstruction
work
used quadratically regularized cost functions
leading to
optimization problems of the form:
\be
\xh =
\argmin{\x \in \complex^N}
\Half \normii{\A \x - \y}^2 + \reg \normii{\T \x}^2
,\ee{e,quad}
where
$\reg > 0$ denotes a regularization parameter
and \T denotes a $K \times N$ matrix transform
such as finite differences.
\iflong{
Some methods based on annihilation filters
also have this form
\cite{ongie:16:otg}.
}
The \term{conjugate gradient (CG) algorithm}
is well-suited to such quadratic cost functions
\cite{pruessmann:01:ais,sutton:03:fii}.
The Hessian matrix
$\A'\A + \reg \T'\T$
often is approximately Toeplitz
\citesl{wajer:01:mso}{,fessler:05:tbi},
so CG with circulant preconditioning
is particularly effective%
\citl{chan:96:cgm}.
Although the quadratically regularized least-squares cost function
\eref{e,quad}
is pass\'e in the compressed sensing era,
CG is often
an inner step
for optimizing
more complicated cost functions%
\citenew{aelterman:11:alb}%
\citl{ramani:11:pmi}.

\subsection{Edge-preserving regularization}

The drawback of the quadratically regularized
cost function
\eref{e,quad}
with \T as finite differences
is that it blurs image edges.
To avoid this blur,
one can replace
the quadratic regularizer
$ \normii{\T \x}^2 $
with a nonquadratic function
\pot{\T \x}
where typically \pot is convex and smooth,
such as the Huber function%
\citl{boubertakh:06:nqc},
a hyperbola%
\citl{husse:04:efo,florescu:14:amm},
or
the Fair potential function
\(
\pot(z) = \del^2 \paren{\abs{z/\del} - \log(1 + \abs{z/\del})}
,\)
among others
\iflong{%
\cite[Ch.~2]{fessler::ir}
}
as follows:
\be
\xh =
\argmin{\x \in \complex^N}
\Kx \defequ
\Half \normii{\A \x - \y}^2 + \reg \pot{\T \x}
.\ee{e,ep}
Such methods have their roots
in Bayesian methods
based on Markov random fields%
\citl{geman:84:srg,besag:86:ots}.
The \term{nonlinear CG algorithm}
is an effective optimization method
for cost functions
with such smooth edge-preserving regularizers.
\iflong{
An interesting alternative
is the complex-valued
\term{3MG (majorize-minimize memory gradient)} algorithm
\cite{florescu:14:amm}.
}
Another appropriate optimization algorithm
is the \term{optimized gradient method (OGM)}%
\citl{kim:16:ofo},
a first-order method
having optimal worst-case performance
among all first-order algorithms
for convex cost functions
with Lipschitz continuous gradients%
\citenew{drori:17:tei}.
OGM has a convergence rate bound
that is twice better than that of
\term{Nesterov's fast gradient method}%
\citenew{nesterov:83:amo}.
\iflong{
A recent \term{line-search OGM} variant
is even more attractive
\cite{drori:18:efo}.
}

\fref{fig,ep}
compares two of these methods
for the case where \T is finite differences
and
\pot is the Fair potential with
$\del = 0.1$,
which approximates TV fairly closely
while being smooth.


\begin{figure*}
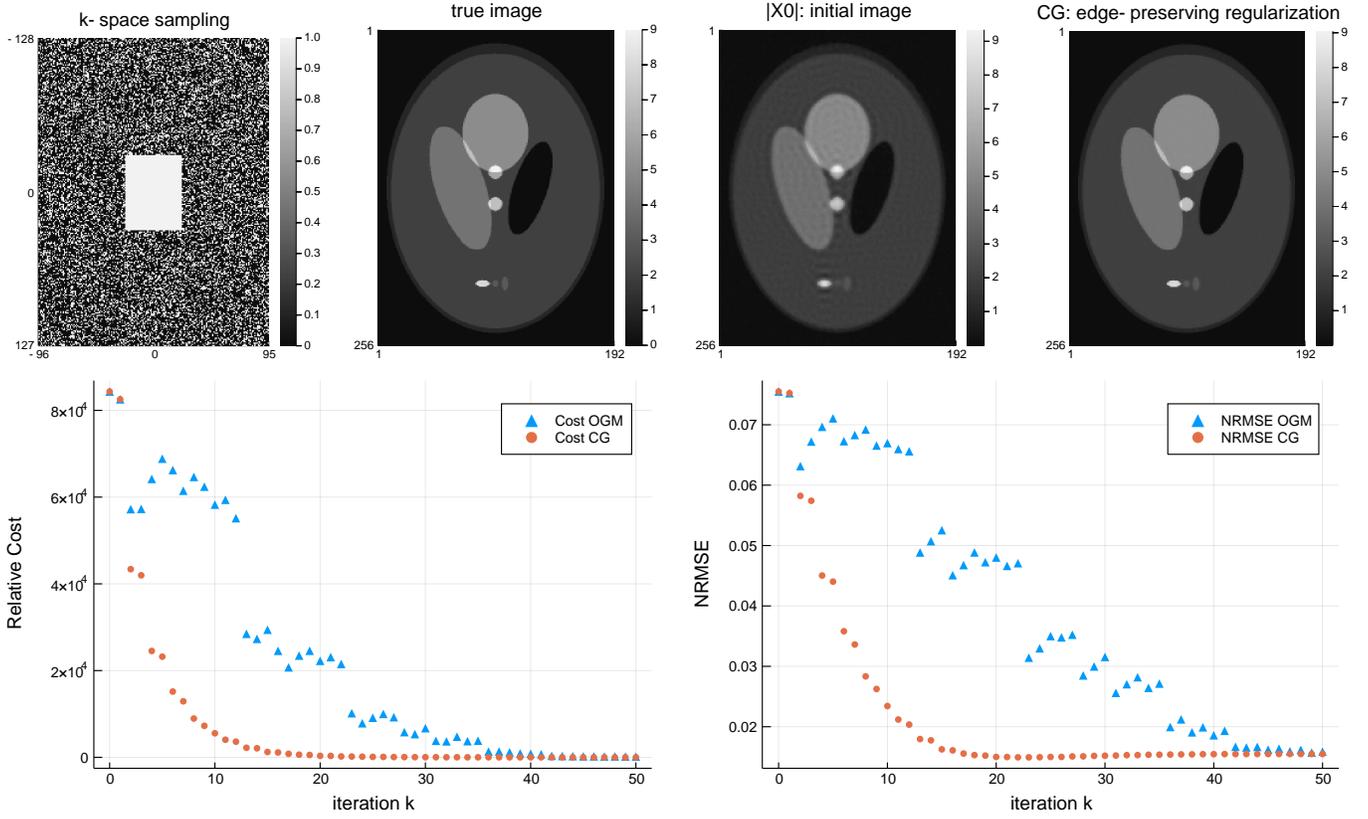

\inceps{fig}{samp}{0.24\linewidth}
\inceps{fig}{xtrue}{0.24\linewidth}
\inceps{fig}{xinit}{0.24\linewidth}
\inceps{fig}{xcg}{0.245\linewidth}
\\
\inceps{fig}{cost_ogm_cg}{0.49\linewidth}
\inceps{fig}{nrmse_ogm_cg}{0.49\linewidth}

\caption{
Comparison of CG and OGM
\add{convergence}
for single-coil MRI reconstruction
with edge-preserving regularization
(akin to anisotropic TV with corner rounding).
From left to right:
Top row:
k-space sampling pattern
where only 34\% of the phase-encodes are collected,
true image,
initial image from zero-filled k-space data,
minimizer \xh
of \eref{e,ep}.
\add{(Both CG and OGM converge to the same limit \xh.)}
Bottom row:
cost function \kost{\x_k} in \eref{e,ep}
and
\add{normalized root mean squared error}
(NRMSE)
$\normii{\x_k - \x} / \normii{\x}$
versus iteration $k$.
}
\label{fig,ep}
\end{figure*}

\subsection{Sparsity models: synthesis form}

Scan time in MRI is proportional
to the number of k-space samples recorded.
Reducing scan time in MRI can reduce cost,
improve patient comfort,
and reduce motion artifacts.
Reducing the number of k-space samples
$ML$ to well below $N$,
necessitates stronger modeling assumptions about \x,
and sparsity models are
prevalent
\citenew{lustig:07:smt}
\citenew{lustig:08:csm}.
Two main categories of sparsity models
are the synthesis approach
and the analysis approach.
In a synthesis model,
one assumes
$\x = \B \z$
for some $N \times K$ matrix \B
where coefficient vector
$\z \in \complex^K$
should be sparse.
In an analysis model,
one assumes
$\T \x$ is sparse,
for some $K \times N$ transformation matrix \T.

A typical cost function
for a synthesis model is
\be
\xh = \B \zh
,\quad
\zh = \argmin{\z \in \complex^K}
\Half \normii{\A \B \z - \y}^2 + \reg \normi{\z}
,\ee{e,synth,l1}
where the 1-norm
is a convex relaxation
of the $\ell_0$ counting measure
that encourages \z to be sparse.
\add{
Typically \B is a wide matrix
(often called an over-complete dictionary)
so that one can represent \x well
using only a fraction of the columns of \B.
}
\iflong{
The optimization formulation
\eref{e,synth,l1}
is also known as the LASSO problem
\cite{tibshirani:96:rsa,candes:09:nim}
and there are numerous algorithms
for solving it.
}
The classical approach
for \eref{e,synth,l1}
is the
\term{iterative soft thresholding algorithm} (ISTA)%
\citenew{daubechies:04:ait},
also known as the
\term{proximal gradient method} (PGM)
\citenew{combettes:11:psm}
and
proximal forward-backward splitting%
\citenew{combettes:05:srb},
having the simple form
\be
\z_{k+1} = \text{soft}\of{\z_k - \D\inv \B'\A'(\A\B \z_k - \y), \reg/\vd}
,\ee{e,ista}
where
the soft thresholding function is defined by
\(
\text{soft}(z,c) = \sign(z) \max(\abs{z} - c, 0)
\)
and
$\D = \diag{\vd}$
is any positive definite diagonal matrix such that
\(
\D - \B'\A'\A\B
\)
is positive semidefinite%
\cite{muckley:15:fpm}.

\add{
The ISTA update \eref{e,ista}
applies to the 1-norm in
\eref{e,synth,l1}.
If we replace that 1-norm with
some other function $\pot(\z)$,
then
one replaces \eref{e,ista}
with the more general
PGM update of the form
\[
\z_{k+1} = \text{prox}_{\reg \pot}
\of{\z_k - \D\inv \B'\A'(\A\B \z_k - \y), \reg/\vd}
,\]
where the \term{proximal operator}
is defined by
\[
\prox_f(\vv) \defequ \argmin{\x} \Half \normii{\x - \vv}^2 + f(\x)
.\]
}

Traditionally
\(
\D = \mnormii{\B'\A'\A\B} \I
,\)
but computing that spectral norm
(via the power iteration)
requires considerable computation
for parallel MRI problems
in general.
However, for Cartesian sampling,
$\F'\F \mleq N \I$
so it suffices to have
\(
N \B'\C'\C\B \mleq \D
.\)
Often the sensitivity maps
are normalized such that
\(
\C'\C = \I
\)
in which case
\(
N \B'\B \mleq \D
\)
suffices.
If in addition
\B^' is a Parseval tight frame,
then
\(
\B'\B \mleq \I
\)
so using
$ \D = N \I$ is appropriate.
For non-Cartesian sampling,
or non-normalized sensitivity maps,
or general choices of \B,
finding \D is more complicated%
\cite{muckley:15:fpm}.

Although ISTA is simple,
it has an undesirably slow
$O(1/k)$ convergence bound,
\add{
where $k$ denotes the number of iterations.}
This limitation was first overcome
by the
\term{fast iterative soft thresholding algorithm} (FISTA)
\citesl{beck:09:fgb}{,beck:09:afi},
also known as the
\term{fast proximal gradient method} (FPGM)
that has an
$O(1/k^2)$ convergence bound.
A recent extension
\iflong{
of this line of proximal methods
}
is the
\term{proximal optimized gradient method} (POGM)
that has worst-case convergence bound
about twice better
than that of FISTA/FPGM%
\citesl{kim:18:aro}{,taylor:17:ewc-composite}.
Both FISTA and POGM
are essentially as simple to implement
as \eref{e,ista}.
Recent MRI studies
have shown POGM converging faster than FISTA,
as one would expect based on the convergence bounds%
\citesl{elgueddari:18:scn-sam}{,lin:19:edp},
particularly when combined with adaptive restart
\citesl{kim:18:aro}{,odonoghue:15:arf}.
So POGM (with restart) is a recommended method
for optimization problems
having the form
\eref{e,synth,l1}.
\iflong{
This topic remains an active research area
with new variants of FISTA appearing recently
\cite{liang:18:iff}.
}
Table~\ref{t,pogm}
provides POGM pseudo-code
for solving composite optimization problems
like the MRI synthesis reconstruction model
\eref{e,synth,l1}.

\fref{fig,odwt}
shows that POGM converges faster than FISTA and ISTA
for minimizing
\eref{e,synth,l1}.


\iflong{
\begin{figure*}
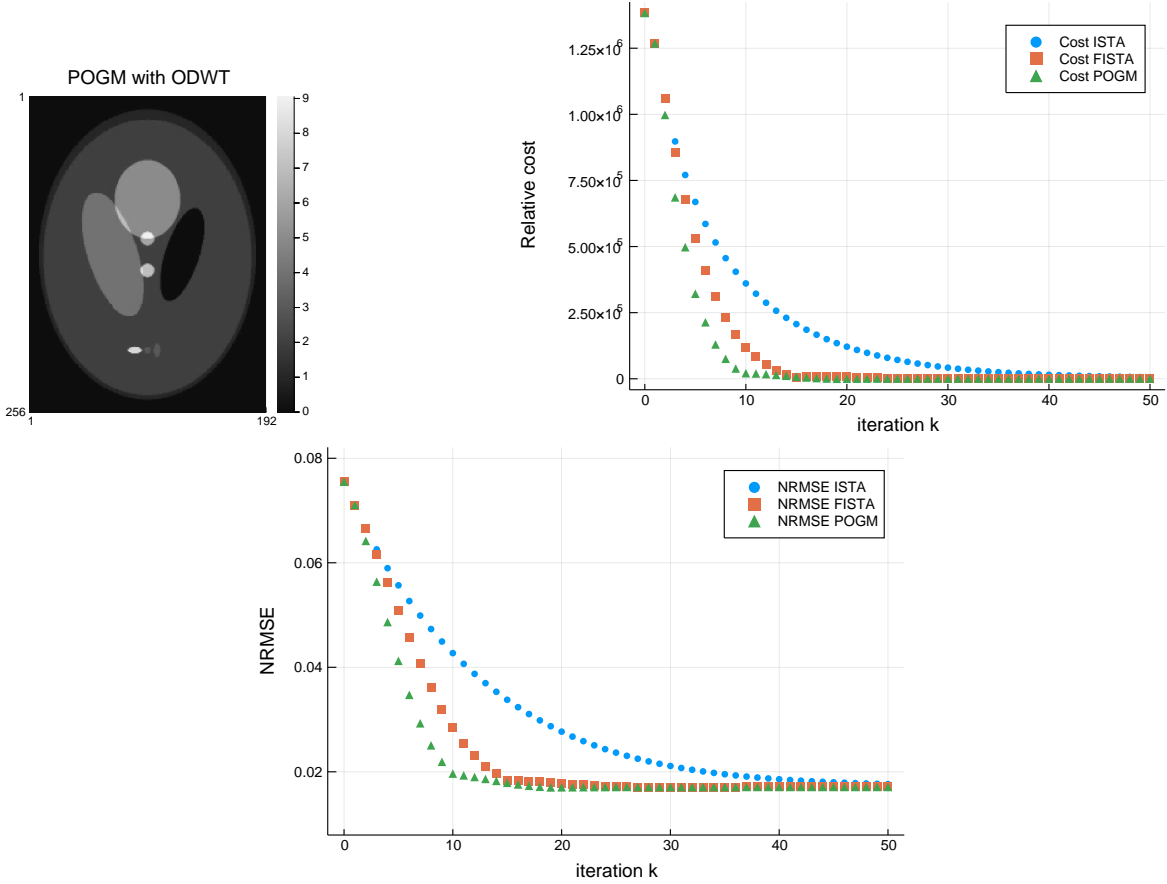

}
\ifshort{
\begin{figure}
}
\inceps{fig}{xpogm_odwt}{0.24\textwidth}
\iflong{
\inceps{fig}{cost_pogm_odwt}{0.49\textwidth}
\\
}
\inceps{fig}{nrmse_pogm_odwt}{0.49\textwidth}

\caption{
Comparison of ISTA/\add{PGM}, FISTA/FPGM and POGM
for single-coil MRI reconstruction
with orthogonal discrete wavelet transform
sparsity regularizer using the 1-norm.
%
Minimizer \xh
of \eref{e,synth,l1};
\iflong{
cost function 
for \eref{e,synth,l1};
}
NRMSE
versus iteration $k$.
FISTA requires about 40\% more iterations to converge than POGM,
consistent with the $2\times$ better worst-case bound of POGM. 
}
\label{fig,odwt}
\iflong{
\end{figure*}
}
\ifshort{
\end{figure}
}

\begin{figure}

\mbox{}\hrulefill\mbox{}

Initialize
$
\w_0 = \x_0 
,$
$\theta_0 = 1
.$
Then
\texttt{for} $k=1:N$:
\desa{
\theta_k &= \leftbrace{
\Half \paren{1 + \sqrt{4 \theta_{k-1}^2 + 1}}, & k < N
\\
\Half \paren{1 + \sqrt{8 \theta_{k-1}^2 + 1}}, & k = N
}
\\
\gam_k &= \frac{1}{L}
\frac{2 \theta_{k-1} + \theta_k - 1}{\theta_k}
}
\desa{
\w_k &= \x_{k-1} - \rat 1/L \nabla f(\x_{k-1})
\\
\z_k &= \w_k
+ \frac{\theta_{k-1} - 1}{\theta_k} (\w_k - \w_{k-1})
+ \frac{\theta_{k-1}}{\theta_k} (\w_k - \x_{k-1})
\\& \hspace*{2em}
+ \frac{\theta_{k-1} - 1}{L \gam_{k-1} \theta_k} (\z_{k-1} - \x_{k-1})
\\
\x_k &= \prox_{\gam_k g}(\z_k)
= \argmin{\x} \Half \normii{\x - \z_k}^2 + \gam_k g(\x)
}

\mbox{}\hrulefill\mbox{}

\caption{
POGM method
\citl{taylor:17:ewc-composite}
for minimizing
\(
f(\x) + g(\x)
\)
where $f$ is convex
with $L$-Lipschitz smooth gradient
and $g$ is convex.
See
\cite{kim:18:aro}
for adaptive restart version.
}
\label{t,pogm}
\end{figure}

\subsection{Sparsity models: analysis form}

A potential drawback
of the synthesis formulation
\eref{e,synth,l1}
is that
\(
\x \approx \B \z
\)
may be a more realistic assumption
than the strict equality
\(
\x = \B \z
\)
when \z is sparse.
The analysis approach
avoids constraining \xh
to lie in any such subspace
(\add{or union of subspaces when \B is wide}).
For an analysis form sparsity model,
a typical optimization problem
involves a composite cost function
consisting of the sum
of a smooth term and a non-smooth term:
\be
\xh = \argmin{\x} \Half \normii{\A \x - \y}^2 + \reg \normi{\T \x}
,\ee{e,l1,anal}
where \T is a sparsifying operator
such as a wavelet transform,
\add{
or finite differences,
or both
\citenew{lustig:07:smt}.
The expression
\eref{e,l1,anal} is general enough
to handle combinations
of multiple regularizers,
such as wavelets and finite differences
\cite{lustig:08:csm},
by stacking the operators in \T
and possibly allowing a weight 1-norm.
}
When \T is finite differences,
the regularizer is called
total variation (TV)%
\citenew{block:07:urm},
and combinations of TV
and wavelet transforms
are useful
\cite{lustig:08:csm}.
Although the details are proprietary,
the FDA-approved method
for compressed sensing MRI
\add{for at least one manufacturer}
is related to \eref{e,l1,anal}%
\citl{forman:16:csa}
\citenew{wetzl:17:hrd}.

\iflong{
We write an ordinary 1-norm in \eref{e,l1,anal},
but some clinical MRI scanners
use a weighted norm
that regularizes the high-frequency components more
\cite{liu:12:dcm}. 
}

When \T is invertible,
such as an orthogonal wavelet transform,
one rewrites the optimization problem
\eref{e,l1,anal}
as
\[
\xh = \T\inv \zh
,\quad
\zh = \argmin{\z} \Half \normii{\A \T\inv \z - \y}^2 + \reg \normi{\z}
,\]
which is simply a special case
of \eref{e,synth,l1}
with $\B = \T\inv$.
\add{
Typically \B is wide
and \T is tall
so this simplification
is not possible in general.
}

In the general case
\eref{e,l1,anal}
where \T is not invertible,
the optimization problem is much harder
than
\eref{e,synth,l1}
due to the non-differentiability of the 1-norm
with the matrix \T.
\add{
The non-invertible case (with redundant Haar wavelets) is used clinically
\citl{liu:12:dcm,ma:19:a4f}%
\citenew{wetzl:17:hrd}.
}
The PGM for
\eref{e,l1,anal}
is
\be
\x_{k+1} = \argmin{\x} \rat\Lips/2 \normii{ \x - \x_k^~ }^2
+ \reg \normi{\T \x}
,\ee{e,pgm}
where
\(
\x_k^~ \defequ \x_k - \rat1/\Lips \A'(\A \x_k - \y)
\)
denotes the usual gradient update
and the Lipschitz constant is
$ \Lips = \mnormii{\A}^2 $.
Unfortunately
there is no simple solution
for computing the
\add{
proximal operator
(defined after \eref{e,ista} above)}
in \eref{e,pgm} in general,
so inner iterative methods are required,
typically involving dual formulations%
\citl{chambolle:04:aaf}\citenew{beck:09:fgb}.
This challenge
makes PGM and FPGM and POGM
less attractive for
\eref{e,l1,anal}
and
%
has led to a vast literature
on algorithms for problems like
\eref{e,l1,anal},
with no consensus on what is best.
The difficulty of
\eref{e,pgm}
is the main drawback of analysis regularization,
whereas a possible drawback of the synthesis regularization
in
\eref{e,synth,l1}
is that often $K \gg N$
for overcomplete \B.

\subsubsection{Approximate methods}
\mbox{}

One popular ``work around'' option
is to ``round the corner'' of the 1-norm,
making smooth approximations like
\(
\abs{z} \approx \sqrt{\abs{z}^2 + \eps}
.\)
This approximation is simply
the hyperbola function
that has a long history
in the edge-preserving regularization literature.
All of the gradient-based algorithms mentioned
for edge-preserving regularization above
are suitable candidates
when a smooth function
replaces the 1-norm.
Smooth functions can shrink values towards zero,
but their proximal operators
never have a thresholding effect
that induces sparsity
by setting many values
exactly to zero.
Whether a thresholding effect is truly essential
is an open question.
\iflong{
\par
One way to overcome the challenge
of the matrix \T in the 1-norm
in \eref{e,l1,anal}
is to replace
\eref{e,l1,anal}
with the following alternative
\cite{wang:08:ana}:
\dena{
\xh &= \argmin{\x} \Half \normii{\A \x - \y}^2 + \reg R_{\alf}(\x)
\nonum
\\
R_{\alf}(\x) &= \min_{\z} \Half \normii{\T \x - \z}^2
+ \alf \normi{\z}
}{e,split0}
where $\alf > 0$.
At first glance
this formulation
appears to enforce sparsity
due to the presence of the 1-norm.
However,
one can solve for \z
and substitute back in
to show that
\(
R_{\alf}(\x) = \pot(\T\x,\alf)
\)
where \pot is the Huber function
with parameter \alf,
so
\eref{e,split0}
is simply another example
of corner rounding
with an approximate 1-norm.
One can show
$\rat 1/\alf R_{\alf}(\x) \arrow \normi{\T \x}$
as $\alf \arrow 0$.
A drawback of
\eref{e,split0}
is that one must choose
the additional regularization parameter \alf
that can affect both the image quality of \xh
and the convergence rate
of iterative algorithms for
\eref{e,split0}.

Another option
is to use an iterative reweighted least-squares
approach like FOCUSS
\cite{ye:07:prm}
that approaches the 1-norm
in the limit as the number of iterations grows,
but is effectively equivalent
to a corner-rounded 1-norm
for any finite number of iterations.
}
Hereafter we focus
on methods that tackle the 1-norm directly
without any such approximations.

\subsubsection{Variable splitting methods}
\mbox{}

Variable splitting methods
replace
\eref{e,l1,anal}
with an exactly equivalent
constrained minimization problem
involving an auxiliary variable
such as
$\z = \T \x$,
\eg,
\dena{
\xh = \argmin{\x}
\min_{\z \,: \, \z = \T\x}
\Half \normii{\A \x - \y}^2 + \reg \normi{\z}
.}{e,z=Tx}
This approach underlies
the
\term{split Bregman algorithm}%
\cite{goldstein:09:tsb},
various
\term{augmented Lagrangian methods}
\citesl{ramani:11:pmi}{,aelterman:11:alb},
and
the
\term{alternating direction multiplier method}
(ADMM)
\citenew{eckstein:92:otd}.
The
augmented Lagrangian
for
\eref{e,z=Tx}
is
\desa{
L(\x,\z;\vgam,\mu)
&=
\Half \normii{\A \x - \y}^2 + \reg \normi{\z}
\\&
+ \real{\inprod{\vgam}{\T\x - \z}}
+ \rat \mu/2 \normii{\T\x - \z}^2
,}
where $\vgam \in \complex^K$
denotes the vector of Lagrange multipliers%
\iflong{%
\footnote{%
One can think of
$\vgam_R = \real(\vgam)$
and
$\vgam_I = \imag(\vgam)$
as the Lagrange multipliers
for the two constraints
$\real(\T\x - \z) = \vzero$
and
$\imag(\T\x - \z) = \vzero$,
and then note that
\(
\real{\inprod{\vgam}{\T\x - \z}}
= \inprod{\vgam_R}{\real(\T\x - \z)}
+ \inprod{\vgam_I}{\imag(\T\x - \z)}
.\)
}
}
\ and $\mu > 0$
is an AL penalty parameter
that affects the convergence rate
but not the final image \xh.
Defining
the scaled dual variable
\(
\veta \defequ 1/\mu \vgam
\)
and completing the square
leads to the following scaled augmented Lagrangian:
\desa{
L(\x,\z;\veta,\mu)
&=
\Half \normii{\A \x - \y}^2 + \reg \normi{\z}
\\& \hspace*{2em}
+ \rat \mu/2
\paren{ \normii{\T\x - \z + \veta}^2 - \normii{\veta}^2 }
.}
An augmented Lagrangian approach
alternates between
descent updates of the primal variables \x, \z
and an ascent update of the scaled dual variable \veta.
The \z update is simply soft thresholding:
\[
\z_{k+1} = \text{soft}(\T\x_k + \veta_k, \reg/\mu)
.\]
The \x update
minimizes a quadratic function:
\[
\x_{k+1} = (\A'\A + \mu \T'\T)\inv
(\A' \y + \mu \T' (\z_{k+1} + \veta_k))
.\]
A few CG iterations
\iflong{
(with an appropriate preconditioner)
}
is a natural choice
for approximating the \x update.
Finally the \veta update is
\[
\veta_{k+1} = \veta_k + (\T\x_{k+1} - \z_{k+1})
.\]
The unit step size here
ensures dual feasibility
\cite{boyd:10:doa}.
A drawback of variable splitting methods
is the need to select the parameter $\mu$.
Adaptive methods have been proposed
to help with this tuning%
\citenew{boyd:10:doa}%
\citl{xu:17:aaw,wohlberg:17:app}.
\iflong{
The above updates of \x and \z are sequential;
parallel ADMM updates are also possible
\cite{eckstein:94:pad,le:17:ecs}.
}

\add{
One could apply ADMM
to the synthesis regularized problem
\eref{e,synth,l1},
though again it would require parameter tuning
that is unnecessary with POGM.
}

The conventional variable split in
\eref{e,z=Tx}
ignores the specific structure
of the MRI system matrix \A
in \eref{e,y=Ax}.
Important properties of \A
include the fact that
$\F'\F$ is circulant
(for Cartesian sampling)
or Toeplitz
(for non-Cartesian sampling)
and that each coil sensitivity matrix \C_l
is diagonal.
In contrast,
the Gram matrix
$\A'\A$
for parallel MRI
is harder to precondition,
though possible%
\citesl{koolstra:19:acs}{,ong:19:anc}.
An alternative splitting 
that simplifies the updates is
\cite{ramani:11:pmi}:
\dena{
&\argmin{\x \in \complex^N}
\min_{\vu \in \complex^{NL}, \, \z \in \complex^K, \, \vv \in \complex^N}
\Half \normii{\F_L \vu - \y}^2 + \reg \normi{\z}
\nonum
\\&
\text{ sub. to }
\vu = \C \x 
,\
\z = \T \vv 
,\
\vv = \x 
,}{e,alp2}
where
\(
\F_L \defequ \I_L \otimes \F
.\)
With this splitting,
the \z update again is simply soft thresholding,
and the \x update involves the diagonal matrix
$\C'\C$ which is trivial.
The \vv update involves
the matrix
$ \T'\T $
that is circulant for periodic boundary conditions
or is very well suited to a circulant preconditioner
otherwise,
using simple FFT operations.
The \vu update
involves the matrix
$ \F_L'\F_L $
that is circulant or Toeplitz. 
This approach exploits
the structure of \A
to simplify the updates;
the primary drawback
is that it requires selecting
even more AL penalty parameters;
condition number criteria
can be helpful
\cite{ramani:11:pmi}.
\iflong{
Many variations are possible,
such as exploiting the fact
that
$\T'\T$
has block tridiagonal structure
when \T involves finite differences
\cite{le:17:ecs}.
}
Another splitting with fewer auxiliary variables
leads to an inner update step
that requires solving denoising problems
similar to \eref{e,pgm}%
\citl{chen:12:faf}.

\subsubsection{Primal-dual methods}
\mbox{}

A key idea behind duality-based methods
is the fact:
\[
\normi{\T \x}
= \max_{\z \in \complex^K \, : \, \norminf{\z} \leq 1}
\real{\inprod{\z}{\T\x}}
.\]
Thus the
(nonsmooth) analysis regularized problem
\eref{e,l1,anal}
is equivalent to this constrained problem:
\be
\argmin{\x} \min_{\z \in \cZ}
\Half \normii{\A \x - \y}^2 + \reg \real{\inprod{\z}{\T\x}}
,\ee{e,l1,anal,dual}
where
\(
\cZ \defequ \cset{\z \in \complex^K}{\norminf{\z} \leq 1}
.\)
The \term{primal-dual methods} 
typically
alternate between updating
the primal variable \x
and the dual variable \z,
using more convenient alternatives to
\eref{e,l1,anal,dual}
that involve separate multiplication
by \A and by $\A'$
without requiring inner CG iterations.
These methods
provide convergence guarantees
and acceleration techniques
that lead to
$O(1/k^2)$ rates%
\citenew{chambolle:11:afo}
\citl{pock:11:dpf,
chen:12:faf,
ong:19:anc,
combettes:12:pds,
condat:13:apd,
sidky:12:cop,
vu:13:asa,
valkonen:14:apd}.
A drawback of such methods
is they typically require
power iterations
to find a Lipschitz constant,
and, like AL methods,
have tuning parameters
that affect the practical convergence rates.
Finding a simple, convergent, and tuning-free method
for the analysis regularized problem
\eref{e,l1,anal}
remains an important open problem.

\codo{
compare ADMM,
MFISTA line search
\cite{zibetti:19:mfw},
and
primal-dual methods
}

\subsection{Patch-based sparsity models}

Using \eref{e,l1,anal}
with a finite-difference regularizer
\iflong{
\(
R(\x) = \normi{\T \x}
\)
}
is essentially equivalent
to using patches of size $2 \times 1$.
It is plausible that one can regularize better
by considering larger patches
that provide more context
for distinguishing signal from noise.
There are two primary modes
of patch-based regularization:
synthesis models
and analysis methods.

A typical synthesis approach
attempts to represent each patch
using a sparse linear combination
of atoms from some signal patch dictionary.
Let $\P_p$ denote the $d \times N$ matrix
that extracts the $p$th of $P$ patches
(having $d$ pixels)
when multiplied by an image vector \x.
Then the synthesis model is that
\(
\P_p \x \approx \D \z_p
\)
where \D is a $d \times J$ dictionary,
\add{
such as the discrete cosine transform (DCT)
\citenew{wang:18:ljs},}
and
$\z_p \in \complex^J$
is a sparse coefficient vector
for the $p$th patch.
Under this model,
a natural regularizer is
\be
R(\x) = \min_{\bset{\z_p}} \sum_{p=1}^P \Half \normii{\P_p \x - \D \z_p}^2
+ \alf \normi{\z_p}
.\ee{e,R,D}
\add{
See \cite{wang:18:ljs}
for an extension
to the case of multiple images.
}
The regularizer has an inner minimization
over the sparse coefficients \bset{\z_p},
so the overall problem involves
both optimizing the image \x
and those coefficients.
This structure lends itself
to \term{alternating minimization} algorithms.
\add{
The work in
\cite{wang:18:ljs}
used ISTA for updating $\z_p$;
the results in
\fref{fig,odwt}
suggest that POGM may be beneficial. 
}

A typical analysis approach
for patches
assumes there is a sparsifying transform \bOm
such that
\(
\bOm \P_p \x
\)
tend to be sparse.
\add{
For example,
\citenew{qu:12:umr}
uses a directional wavelet transform
for each patch.}
Under this model,
a natural regularizer is
\be
R(\x) = \min_{\bset{\z_p}} \sum_{p=1}^P \Half \normii{\bOm \P_p \x - \z_p}^2
+ \alf \normi{\z_p}
.\ee{e,R,Om}
Again a double minimization
over the image \x
and the transform coefficients \bset{\z_p} is needed,
so \term{alternating minimization} algorithms
are natural.
\comment{ 
The analysis regularizer
\eref{e,R,Om}
is jointly convex in \x and \bset{\z_p},
whereas the synthesis regularizer
\eref{e,R,D}
is not
due to the product $\D \z_p$.
}
For alternating minimization
(block coordinate descent),
the update of each \z_p is simply soft thresholding,
and the update of \x
is a quadratic problem
involving
$ \A'\A + \reg \sum_p \P_p' \bOm' \bOm \P_p $.
When the transform \bOm is unitary
and the patches are selected
with periodic boundary conditions
and a stride of one pixel,
then this simplifies to
$ \A'\A + \reg \I $.
A few inner iterations
of the (preconditioned) CG algorithm
is useful for the \x update.
Under these assumptions,
and using just a single gradient descent update
for \x,
an alternating minimization algorithm
for least-squares with regularizer
\eref{e,R,Om}
simply alternates between
a denoising step
and a 
gradient step:
\dena{
\x_k^~ &=
\textstyle \sum_{p=1}^P \P_p' \bOm'
\, \text{soft}\of{\bOm \P_p \x_k, \alf}
\label{e,up,Om,xk,denoise}
\\
\x_{k+1} &= \x_k -
(\D + \reg \I)\inv
\paren{\A'(\A\x - \y) + \reg \x_k^~}
\nonum
.}{e,nada}
For this algorithm
the cost function
is monotonically nonincreasing.

\subsection{Adaptive regularization}

The patch dictionary \D
in \eref{e,R,D}
or the sparsifying transform \bOm
in \eref{e,R,Om}
can be chosen
based on mathematical models
like the 
DCT,
or they can be
learned from a population of preexisting training data
and then used
in \eref{e,R,D}
or \eref{e,R,Om}
for subsequent patients.
A third possibility
is to adapt \D or \bOm
to each specific patient%
\citesl{ravishankar:11:mir}{,ravishankar:15:ebc}.
The ``dictionary learning MRI'' (DLMRI) approach
\citenew{ravishankar:11:mir}
uses a \add{non-convex} regularizer of the following form:
\be
R(\x) = \min_{\D \in \cD} \min_{\bset{\z_p}}
\sum_{p=1}^P \normii{\P_p \x - \D \z_p}^2
+ \alf \normi{\z_p}
,\ee{e,dlmri}
where \cD is the feasible set of dictionaries
(typically constrained so that each atom has unit norm).
Now there are three set of variables to optimize:
\x, \bset{\z_p}, \D,
so alternating minimization methods
are well suited.
The update of the image \x
is a quadratic optimization subproblem,
the \z_p update is soft thresholding,
and the \D update
is simple when considering one atom at a time%
\citl{ravishankar:17:eso}.
\iflong{
This problem is nonconvex
because of the $\D \z_p$ product,
but there is some convergence theory for it
\cite{ravishankar:17:eso}.
}

The ``transform learning MRI'' (TLMRI) approach%
\citl{ravishankar:15:ebc}
uses a regularizer of this form:
\[
R(\x) = \min_{\bOm} \min_{\bset{\z_p}} \sum_{p=1}^P
\normii{\bOm \P_p \x - \z_p}^2
+ \alf \normi{\z_p} + \gam r(\bOm)
,\]
where
$r(\bOm)$ enforces or encourages properties
of the sparsifying transform
such as orthogonality.
Again, alternating minimization methods
are well suited;
the \bOm update involves
(small) SVD operations.
\iflong{
See
\cite{ravishankar:16:ddl}
for
convergence theory
and an extension
to learning a union of sparsifying transforms.
}

\codo{compare TV+wavelet regularizer to TLMRI}

\subsection{Convolutional regularizers}

An alternative to patch-based regularization
is to use convolutional sparsity models
\citesl{wohlberg:16:eaf}{,chun:18:cdl,chun:18:cao-asilomar,nguyenduc:18:csd}.
A convolutional synthesis regularizer
replaces
\eref{e,R,D}
with
\[
R(\x) = \min_{\bset{\z_k}}
\Half \normii{\x - \textstyle \sum_{k=1}^K \vh_k * \z_k}^2
+ \alf \normi{\z_k}
,\]
where \bset{\vh_k}
is a set of filters
learned from training images
\cite{wohlberg:16:eaf}
\iflong{
(or from k-space data \cite{ong:18:ksa}) 
}
and $*$ denotes convolution.
Again,
\term{alternating minimization} algorithms
are a natural choice
because the \x update is quadratic
and the \z_k update is a sparse coding problem
for which proximal methods
like POGM are well-suited%
\citl{zhang::acs}.

A convolution analysis regularizer
replaces
\eref{e,R,Om}
with
\[
R(\x) = \min_{\bset{\z_k}} \sum_{k=1}^K
\Half \normii{\vh_k \conv \x - \z_k}^2
+ \alf \normi{\z_k}
.\]
Again,
\term{alternating minimization} algorithms
are effective,
where the \z_k update is soft thresholding.
One can either learn the filters \bset{\vh_k}
from good quality
(\eg, fully sampled) training data,
or adapt the filters
for each patient
by jointly optimizing
\x, \bset{\vh_k} and \bset{\z_k}
using alternating minimization.
\iflong{
For such adaptive regularizers,
constraints on the filters
are essential
\cite{chun:18:cdl,chun:18:cao-asilomar}.
}

\subsection{Other methods}

The summation in
\eref{e,up,Om,xk,denoise}
is a particular type of patch-based denoising
of the current image estimate \x_k.
There are many other denoising methods,
some of which have variational formulations
well-suited to inverse problems,
but many of which do not,
such as
nonlocal means (NLM)%
\citenew{buades:10:idm}
and block-matching 3D (BM3D)%
\citenew{dabov:07:idb}.
One way to adapt most such denoising methods
for image reconstruction
is to use a plug-and-play ADMM approach%
\citesl{chan:17:pap}{,buzzard:18:pap}
that replaces a denoising step like
\eref{e,up,Om,xk,denoise}
\iflong{
that originated from an optimization formulation
}
with a general denoising procedure.
\iflong{
See also
\cite{romano:17:tle}.
}

\subsection{Non-SENSE methods}

The measurement model
\eref{e,s=FCx}
and
\eref{e,y=Ax}
has a single latent image \x,
viewed by each receive coil.
An alternate formulation
is to define
a latent image for each coil
\(
\x_l \defequ \C_l \x
\)
and write the measurement model
as
\(
\y_l = \F \x_l + \vveps_l
.\)
For such formulations,
the problem becomes
to reconstruct the $L$ images
$\X = [\x_1 \ \ldots \ \x_L]$
from the measurements,
while considering relationships
between those images.
Because multiplication
by the smooth sensitivity map \C_l
in the image domain
corresponds to convolution
with a small kernel in the frequency domain,
any point in k-space
can be approximated
by a linear combination of its neighbors
in all coil data
\cite{griswold:02:gap}.
This ``GRAPPA modeling''
leads to an approximate consistency condition
\(
\vecop{\X} \approx \G \vecop{\X}
\)
where \G is a matrix involving small k-space kernels
that are learned from calibration data
\cite{griswold:02:gap}.
This relationship leads
to ``SPIRiT''
\citl{lustig:10:sis}
optimization problems like:
\desa{
\Xh &= \argmin{\X \in \complex^{N \times L}}
\Half \mnormfrob{\F \X - \Y}^2
\\& \hspace*{2em}
+ \reg_1 \Half \normii{(\G - \I) \vecop{\X}}^2
+ \reg_2 R(\X)
,}
where
\(
\Y =
[\y_1 \ \ldots \ \y_L]
\in \complex^{M \times L}
\)
and $R(\X)$ is a regularizer
that encourages joint sparsity
because all of the images \bset{\x_l}
have edges in the same locations%
\citl{murphy:12:fes}.
No sensitivity maps \C
are needed for this approach.
When $\reg_2 = 0$ the problem is quadratic
and CG is well suited%
\citl{lustig:10:sis}.
Otherwise,
ADMM is convenient
for splitting this optimization problem
into parts with easier updates%
\citl{weller:14:alw,duan:18:eos}.
\iflong{
See
\citesl{trzasko:11:cpm,majumdar:12:clm,uecker:14:eae,shin:14:cpi,balachandrasekaran:18:cfb}{,elgueddari:19:cob} 
for subspace and joint sparsity approaches
that go further
by circumventing finding the calibration matrix \G.
}
\add{The ESPIRiT approach
uses the redundancy
in k-space data from multiple coils
to estimate sensitivity maps
from the eigenvectors
of a certain block-Hankel matrix
\cite{uecker:14:eae};
this approach helps bridge
the SENSE and GRAPPA approaches
while building on related signal processing
tools like subspace estimation%
\citl{krim:96:tdo}
and multichannel blind deconvolution%
\citl{harikumar:99:pbr,sharif:11:gfo}.
}

\section{Summary}
Although the title of this paper
is ``optimization methods for...''
before selecting an optimization algorithm
it is far more important
(for under-sampled problems)
to first select
an appropriate cost function
that captures
useful prior information
about the latent object \x.
The literature is replete
with numerous candidate models,
each of which often lead
to different optimization methods.
Nevertheless,
common ingredients
arise in most formulations,
such as alternating minimization
(block coordinate descent)
at the outer level,
preconditioned CG
for inner iterations
related to quadratic terms,
and soft thresholding
or other proximal operators
for nonsmooth terms
that promote sparsity.

This survey has focused on 1-norm regularizers
for simplicity,
but (nonconvex) $p$ ``norms'' with $0 \leq p < 1$
have also been investigated
and appear to be beneficial
particularly for very undersampled measurements%
\citl{trzasko:09:hum}.
This survey considers a single image \x
but many MRI scan protocols
involve several images
with different contrast
and it may be useful
to reconstruct them jointly,
\eg,
by considering common sparsity
or subspace models%
\citl{doneva:10:csr}%
\citl{velikina:13:amp,davies:14:acs,
zhao:15:amp,lee:16:aom,asslander:18:lra,
zhao:18:imr,zibetti:18:a3t}.

There are many open problems in optimization
that are relevant to MRI.
The analysis form regularized problem
\eref{e,l1,anal}
remains challenging,
and further investigation
of analysis vs synthesis approaches
is needed%
\citl{cherkaoui:18:avs}.
There has been considerable recent progress
on finding optimal worst-case methods
\citesl{drori:17:tei}{,kim:16:ofo,drori:18:efo},
but these optimality results
are for very broad classes of cost functions,
whereas the cost functions in MRI reconstruction
have particular structure.
Finding algorithms with optimal complexity
(fastest possible convergence)
for MRI-type cost functions
would be valuable
both for clinical practice
and for facilitating research.

Finally,
the current trend is
to use convolutional neural network (CNN) methods
to process under-sampled images,
or for direct reconstruction,
or as denoising operators%
\citl{akcakaya::dlm,cheng::csf,liang::ifm}.
\iflong{
(Finding stable approaches is crucial
\cite{antun:19:oio}.)
}
The stochastic gradient descent method
\iflong{
(or a variant
\cite{kingma:14:aam})
}
currently is the universal optimization tool
for training CNN models.
Many ``deep learning'' methods for MRI
are based on network architectures
that are ``unrolled'' versions
of iterative optimization methods like PGM
\citl{
aggarwal:19:mmb,
mardani:18:npg-nips,
lee:18:drl,
hammernik:18:lav,
ravishankar:18:ddt%
}.
Thus, familiarity with ``classical'' optimization methods
for MR image reconstruction
is important
even in the machine learning era.

\comment{
See
\cite{fessler:19:omf-arxiv}
for a more complete list of references.
}

\section{Acknowledgement}
The author thanks the reviewers
for their detailed comments
that improved the paper.





%




\comment{
\begin{IEEEbiographynophoto}{Jeffrey~A.~Fessler}
Jeff Fessler received the BSEE degree from Purdue University
in 1985, the MSEE degree from Stanford University in 1986, and
the M.S. degree in Statistics from Stanford University in 1989.
From 1985 to 1988 he was a National Science Foundation Graduate Fellow
at Stanford, where he earned a Ph.D. in electrical engineering
in 1990.  He has worked at the University of Michigan since then.
From 1991 to 1992 he was a Department of Energy Alexander Hollaender
Post-Doctoral Fellow in the Division of Nuclear Medicine.  From 1993
to 1995 he was an Assistant Professor in Nuclear Medicine and the
Bioengineering Program.  He is now a Professor in the Departments
of Electrical Engineering and Computer Science, Radiology, and
Biomedical Engineering.  He became a Fellow of the IEEE in 2006,
for contributions to the theory and practice of image reconstruction.
He received the Francois Erbsmann award for his IPMI93 presentation,
and received Edward Hoffman Medical Imaging Scientist Award in 2013.
He has served as an associate editor
for the IEEE Transactions on Medical Imaging,
the IEEE Transactions on Image Processing,
and the IEEE Signal Processing Letters,
and currently serves as an associate editor
for the IEEE Transactions on Computational Imaging.
He has chaired the IEEE T-MI Steering Committee
and the ISBI Steering Committee.
He was co-chair of the 1997 SPIE conference
on Image Reconstruction and Restoration,
technical program co-chair
of the 2002 IEEE International Symposium on Biomedical Imaging (ISBI),
and general chair of ISBI 2007.
His research interests are in statistical aspects of imaging problems,
and he has supervised doctoral research in PET, SPECT, X-ray CT, MRI,
and optical imaging problems.
\end{IEEEbiographynophoto}
}




\end{document}